\documentclass[conference,9pt]{IEEEtran}

\AtBeginDocument{%
  \providecommand\BibTeX{{%
    \normalfont B\kern-0.5em{\scshape i\kern-0.25em b}\kern-0.8em\TeX}}}
\usepackage{cite}

\usepackage{url}
\usepackage{enumitem}

\usepackage[utf8]{inputenc} 
\usepackage[T1]{fontenc}    
\usepackage{url}            
\usepackage{booktabs}       
\usepackage{amsfonts}       
\usepackage{amsmath}
\usepackage{nicefrac}       
\usepackage{microtype}      
\usepackage{xcolor}         
\usepackage{colortbl}
\usepackage{subcaption}
\usepackage{graphicx}
\usepackage{svg}
\usepackage{multirow}
\usepackage{wrapfig}
\usepackage{threeparttable}
\usepackage{tikz}
\usepackage{makecell}
\usepackage{tablefootnote}
\usepackage[hidelinks]{hyperref}
\usepackage{dblfloatfix}

\usepackage{color}
\begin{document}
\pagestyle{plain}
\title{\textsc{Gamora}: \underline{G}raph Le\underline{a}rning based Sy\underline{m}bolic Reas\underline{o}ning for La\underline{r}ge-Scale Boole\underline{a}n Networks}


\author{Nan Wu$^1$, Yingjie Li$^2$, Cong Hao$^3$, Steve Dai$^4$, Cunxi Yu*$^2$, Yuan Xie$^5$
\\$^1$University of California, Santa Barbara, $^2$University of Utah, $^3$Georgia Institute of Technology,\\ $^4$NVIDIA, $^5$Alibaba DAMO Academy\\ 
nanwu@ucsb.edu, yingjie.li@utah.edu, callie.hao@gatech.edu, sdai@nvidia.com, yuanxie@gmail.com\\
*correspondence: cunxi.yu@utah.edu}


\newcommand{\red}[1]{\textcolor{red}{#1}}

\newcommand{\cy}[1]{\textcolor{violet}{#1}}

\maketitle

\begin{abstract}
Reasoning high-level abstractions from bit-blasted Boolean networks (BNs) such as gate-level netlists can significantly benefit functional verification, logic minimization, datapath synthesis, malicious logic identification, etc. 
Mostly, conventional reasoning approaches leverage structural hashing and functional propagation, suffering from limited scalability and inefficient usage of modern computing power.
In response, we propose a novel symbolic reasoning framework exploiting graph neural networks (GNNs) and GPU acceleration to reason high-level functional blocks from gate-level netlists, namely \textsc{Gamora},
which offers {\bf high reasoning performance} w.r.t exact reasoning algorithms, strong {\bf scalability} to BNs with over 33 million nodes, and {\bf generalization capability} from simple to complex designs. 
To further demonstrate the capability of \textsc{Gamora}, we also evaluate its reasoning performance after various technology mapping options, since technology-dependent optimizations are known to make functional reasoning much more challenging.
Experimental results show that (1) \textsc{Gamora} reaches almost 100\% and over 97\% reasoning accuracy for carry-save-array (CSA) and Booth-encoded multipliers, respectively, with up to six orders of magnitude speedups compared to the state-of-the-art implementation in the ABC framework; 
(2) \textsc{Gamora} maintains high reasoning accuracy ($>$92\%) in finding functional modules after complex technology mapping, and we comprehensively analyze the impacts on \textsc{Gamora} reasoning from technology mapping.
\textsc{Gamora} is available at \textcolor{teal}{\url{https://github.com/Yu-Utah/Gamora}}.

\end{abstract}





\section{Introduction}

Reasoning high-level abstractions (e.g., functional blocks) from bit-blasted Boolean networks (BNs) (e.g., unstructured gate-level netlists) has demonstrated its wide applications in improving functional verification efficiency~\cite{ciesielski2019understanding,mahzoon2019revsca} and identifying malicious logics such as detecting hardware trojan and intellectual property infringement usage~\cite{li2019attacking,botero2021hardware}.
In the era of globalization and democratization of integrated circuit (IC) development and fabrication, such reasoning is expected to bring broader impacts on hardware security, which is at the heart of modern computing systems: more than 40 percent of FPGA/ASIC projects are working under safety-critical development process standards or guidelines~\cite{wilson2022}. 

Due to the optimization conducted by RTL synthesis tools, reasoning high-level abstractions such as functional blocks from unstructured or flattened netlists is extremely challenging, since hierarchy and module information is lost during multi-level logic minimization and technology mapping, which is also complicated by functional blocks overlapping and gate sharing. 
The problem goes further due to the explosion in runtime for large-scale BNs.
Conventional reasoning approaches leverage structural analysis and functional propagation. 
Structural approaches either adopt shape hashing based on circuit topology to find structurally similar wires to form word-level abstractions~\cite{li2013wordrev}, or rely on reference libraries to map sub-circuits with reference circuits~\cite{cakir2018reverse}.
Functional approaches focus on identifying functionally equivalent gates and wires by cut enumeration~\cite{subramanyan2013reverse,gascon2014template}.
The combination of structural and functional analysis \cite{li2013wordrev,subramanyan2013reverse,yu2017fast} is more prevalent for efficient word-level abstraction and propagation.
Despite the achieved success, the performance of these conventional approaches is restricted by \textbf{limited scalability} and \textbf{inefficient utilization of modern computing power}:
(1) structural hashing is very time/memory-consuming for large BNs with billions of nodes;
(2) functional propagations by symbolic evaluation are solver-ready but extremely expensive, in particular for \textit{bit-blasted} non-linear arithmetic BNs;
(3) all these algorithms do not effectively utilize modern computing power due to the difficulty of parallelism.

Recently, we have witnessed the emergence of machine learning (ML) applied for computer systems and electronic design automation (EDA) tasks~\cite{wu2022survey}, as an alternative to conventional design solutions.
Since circuit netlists or BNs can be easily represented as graphs, graph neural networks (GNNs) are naturally suitable to classify sub-circuit functionality from gate-level netlists~\cite{alrahis2021gnn}, analyze impacts of circuit rewriting on functional operator detection~\cite{zhao2022graph}, and predict boundaries of arithmetic blocks~\cite{he2021graph}.

\begin{figure}[t]
    \centering
    \vspace{2pt}
    \includegraphics[width=\linewidth]{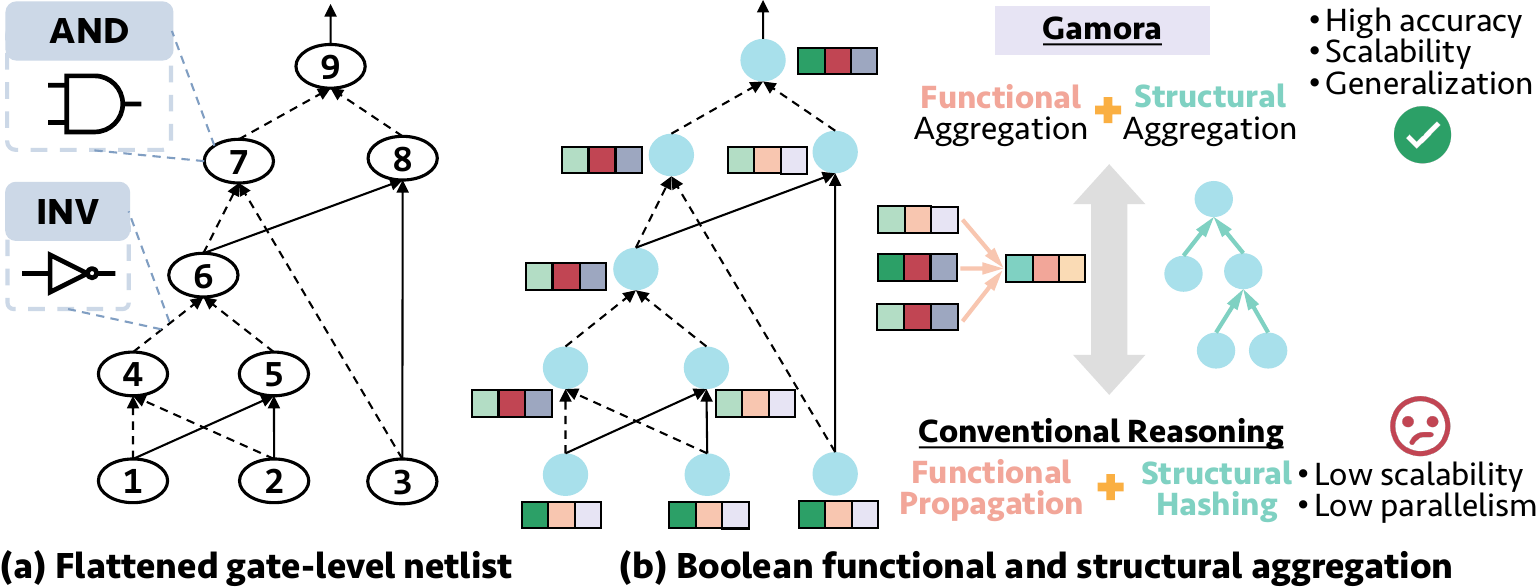}
    \vspace{-10pt}
    \caption{The inputs to \textsc{Gamora} are flattened gate-level netlists, with each node as an AND gate and dashed edges as inverters. By encoding Boolean functional information as node features, \textsc{Gamora} can simultaneously handle functional and structural aggregation, analogous to functional propagation and structural hashing in conventional reasoning but with strong scalability.}
    \vspace{-10pt}
    \label{fig:high-level}
\end{figure}

Motivated by the limitations of conventional approaches and the potentials of GNNs applied on circuit designs, we propose a graph learning-based symbolic reasoning framework to reverse engineer functional blocks from gate-level netlists, namely \textsc{Gamora}, which has \textbf{high reasoning accuracy}, \textbf{strong scalability} to BNs with billions of nodes, and \textbf{generalization capability} from simple to complex designs.
\textsc{Gamora} employs a multi-task GNN to guarantee reasoning accuracy while simultaneously handling structural and functional information from BNs.
Once well trained, \textsc{Gamora} becomes adept at generalizing to large-scale and complex BNs, leveraging the accelerated inference and parallel processing offered by modern computing systems.
We summarize our contributions as follows.


\begin{itemize}[leftmargin=*]
\item \textbf{Novel multi-task GNN for structure and function fusion.} 
The message passing mechanisms in GNNs enable simultaneous \textit{Boolean functional} and \textit{structural aggregation}, corresponding to the symbolic propagation and structural hashing in conventional reasoning methods, as shown in Figure~\ref{fig:high-level}.
The multi-task setting allows knowledge sharing across different reasoning sub-tasks to guarantee high reasoning accuracy.


\item \textbf{Billion-node scalability and parallelism.} 
We develop domain-specific techniques to compress node features, significantly reducing compute costs.
The exploitation of graph learning draws better support from modern computing systems, such as GPU deployment, for scalability to large BNs and parallel execution.


\item {\bf Generalization capability.} 
Unlike many ML-based approaches that are trained with complex designs and infer on simpler ones, \textsc{Gamora} can easily generalize from simple to complex BNs and handle the reasoning complexity introduced by more advanced designs (such as Booth-encoded multipliers) and technology mapping.




\item \textbf{Evaluation.} 
Regarding reasoning performance, \textsc{Gamora} reaches almost 100\% and over 97\% reasoning accuracy for carry-save array (CSA) and Booth multipliers, respectively;
after technology mapping, the reasoning accuracy is still over 92\%.
Regarding runtime and scalability, \textsc{Gamora} can perform reasoning for large BNs with tens of millions of nodes/edges within one second, with a speedup of up to six orders of magnitude compared to the logic synthesis tool ABC~\cite{brayton2010abc}.

\end{itemize}

\section{Preliminary and Motivation}

\subsection{Boolean Networks and And-Inverter Graphs}
BNs are well-studied discrete mathematical models with broad applications in chemistry, biology, circuit design, formal verification, etc.
For purposes of synthesis and verification, a concise and uniform representation of BNs consisting of inverters and two-input AND-gates, known as and-inverter graphs (AIGs), has found successful use in diverse EDA tasks, since AIGs allow rewriting, simulation, technology mapping, placement, and verification to share the same data structure~\cite{mishchenko2006dag}.
In an AIG, each node has at most two incoming edges; 
a node without incoming edges is a primary input (PI);
primary outputs (POs) are denoted by special output nodes;
each internal node represents a two-input AND function. 
Based on De Morgan’s laws, any combinational BN can be converted into an AIG~\cite{brayton2010abc} in a fast and scalable manner.

\begin{figure}[t]
    \centering
    \includegraphics[width=\linewidth]{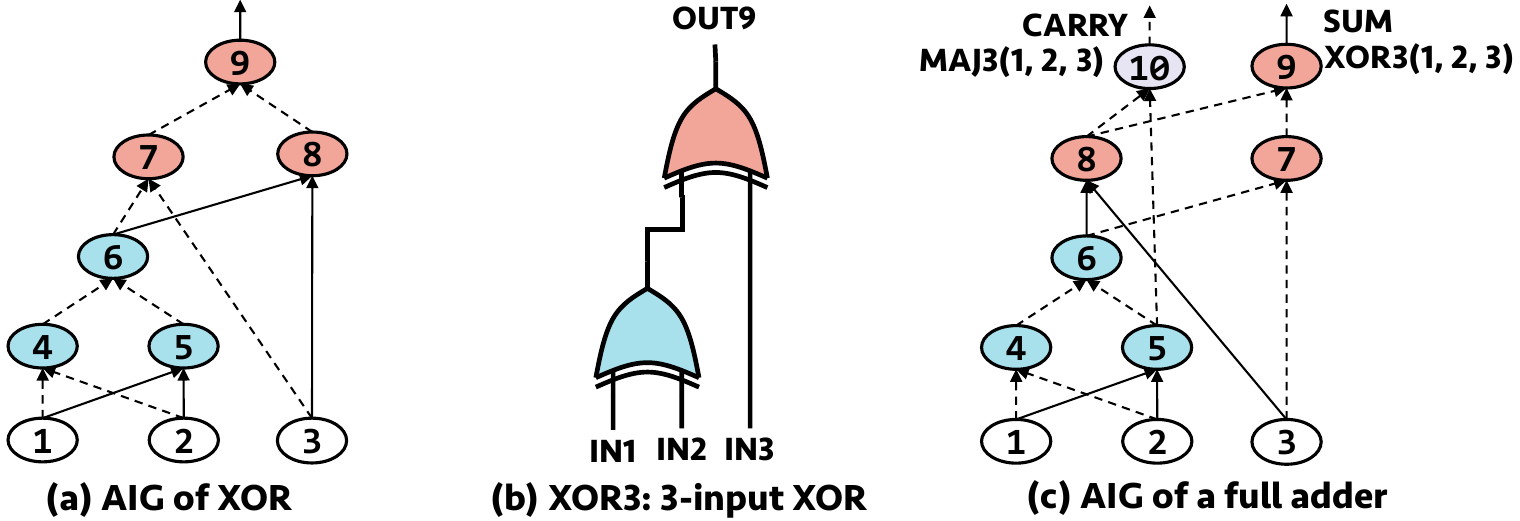}
    \caption{Netlists of XOR and a full adder. (a) AIG of XOR3 function. (b) XOR3 function: OUT9 = XOR3(IN1, IN2, IN3). (c) Full adder with a sum function (i.e., XOR3) and a carry-out function (i.e., MAJ3).}
    \label{fig:adder}
\end{figure}

In AIGs, cut enumeration can be used to detect Boolean functions.
A feasible cut of node $n$ is a set of nodes in the transitive fan-in cone of $n$, whose truth value assignments completely determine the value of $n$.
A cut is $K$-feasible if there are no more than $K$ inputs.
Figure~\ref{fig:adder} depicts an example of reasoning XOR functions and full adders from AIGs.
In Figure~\ref{fig:adder}(a), the AIG has a 3-feasible cut of node 9 and a 2-feasible cut of node 6; after truth table computation, the functions of node 6 and node 9 are IN1$\oplus$IN2 and IN1$\oplus$IN2$\oplus$IN3, respectively.
Thus, as shown in Figure~\ref{fig:adder}(b), node 6 is an XOR2 function, and node 9 is an XOR3 function.
Figure~\ref{fig:adder}(c) shows a full adder bitslice, with the sum as an XOR function and the carry-out as a majority (MAJ) function.
By pairing an XOR3 with a MAJ3 with identical inputs, a full adder bitslice can be extracted, which is then aggregated for word-level abstraction. 



\subsection{Word-Level Abstraction}
Word-level abstraction significantly reduces the complexity of large-scale BNs by grouping wires into meaningful words and keeping useful information related to control logic, which is widely applied in reasoning functional units from gate-level netlists~\cite{li2013wordrev,subramanyan2013reverse,yu2017fast}.
Conventional word identification uses \textit{structural shape hashing} and  \textit{functional bitslice aggregation}. 
Structural shape hashing assigns each edge in the BN a shape, which is defined as the directed graph constructed by the backward reachable nodes from this edge within certain depth/steps.
Functional bitslice aggregation adopts functional matching to group functionally equivalent nodes and edges by cut enumeration.
Typically, structural hashing and functional aggregation are iteratively propagated across neighborhood nodes using symbolic evaluation~\cite{li2013wordrev,subramanyan2013reverse,yu2017fast}.
However, for large-scale BNs, structural hashing is memory-consuming;
functional bitslice aggregation is not efficient due to the requirement of bit-blasting;
the computation of symbolic evaluation is also expensive.
Motivated by the \textbf{limited scalability} and the \textbf{difficulty of parallelism}, we propose to exploit \textbf{graph learning and GPU acceleration for highly scalable reasoning}.

\subsection{Graph Neural Network}
Since BNs and circuit netlists are naturally represented as graphs, GNNs can be leveraged to classify sub-circuit functionality from gate-level netlists~\cite{alrahis2021gnn}, predict the functionality of approximate circuits~\cite{bucher2022appgnn}, analyze impacts of circuit rewriting on functional operator detection~\cite{zhao2022graph}, and predict boundaries of arithmetic blocks~\cite{he2021graph}.
Promising as they are, these approaches focus on graphs with tens of thousands of nodes, and conduct training on complex designs and inference on relatively simpler ones, in which the generalization capability from simple to complex designs is not well examined.

GNNs operate by propagating information along the edges of a given graph.
Each node is initialized with a representation, which could be either a direct representation or a learnable embedding obtained from node features.
Then, a GNN layer updates each node representation by integrating node representations of both itself and its neighbors in the graph.
The propagation along edges extracts structural information from graphs, corresponding to structural shape hashing in conventional reasoning;
after encoding Boolean functionality into node features, neighborhood aggregation is analogous to functional aggregation in conventional reasoning.
Thus, the inherent message-passing mechanism in GNNs enables simultaneously handling structural and functional information.
Motivated by \textbf{the analogy between GNN computation and conventional reasoning}, we propose \textbf{a multi-task GNN for high-performance reasoning} w.r.t. exact reasoning algorithms, with \textbf{strong generalization capability} from simple to complex designs.

\begin{figure*}[t]
    \centering
    \includegraphics[width=\textwidth]{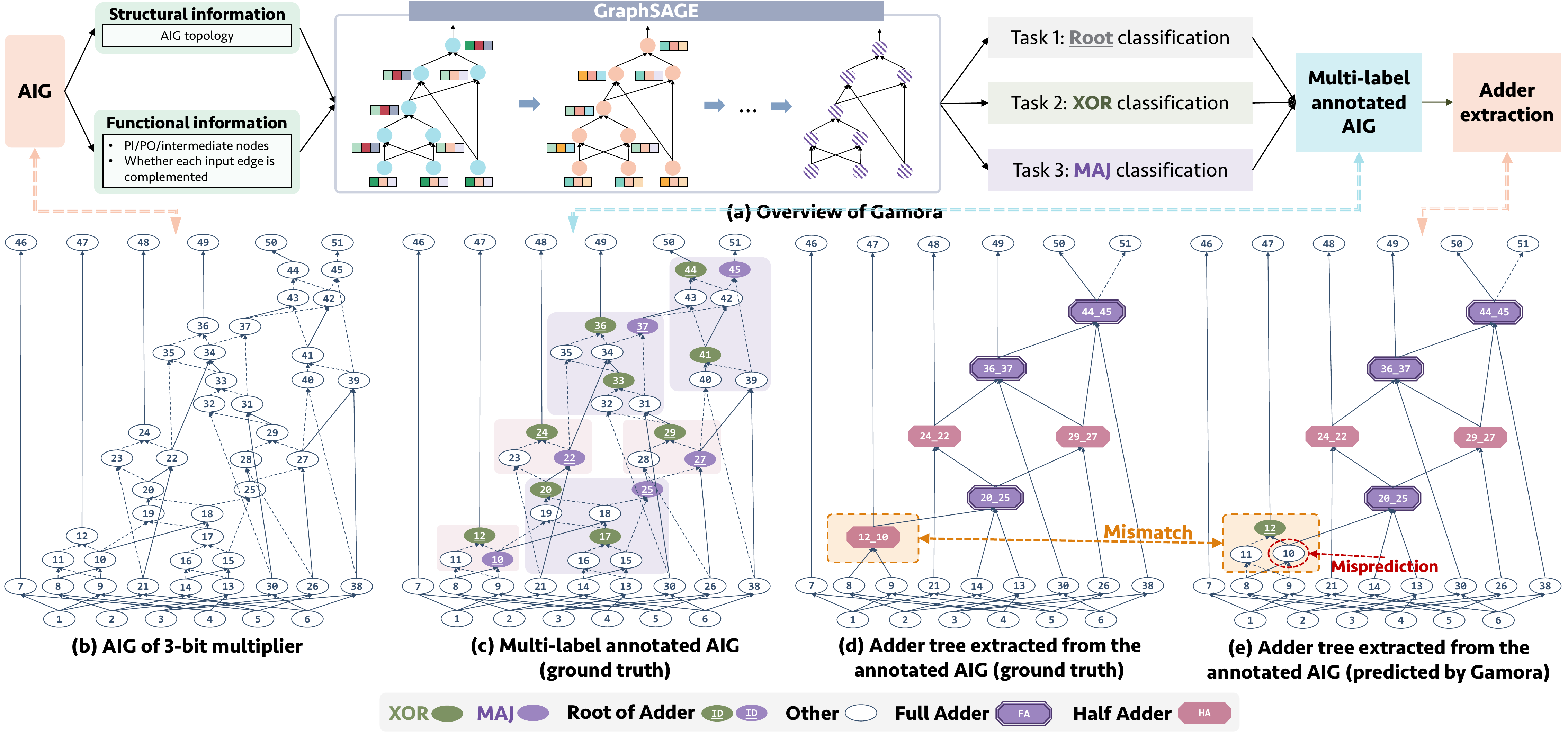}
    \caption{Overview of \textsc{Gamora}. (a) \textsc{Gamora} takes in flattened netlists in AIG format and performs multi-task node classification to reason the Boolean function of each node, after which the adder trees within multiplier netlists can be automatically extracted to improve the efficiency of word-level abstraction.
    (b) AIG of a 3-bit CSA multiplier after synthesis. 
    (c) Annotated AIG with the Boolean function of each node, using the ground truth provided by ABC.
    (d) Adder tree extracted based on the \textit{exact} reasoning, including three FAs and three HAs.
    (e) Adder tree extracted based on the reasoning performed by \textsc{Gamora}.}
    \vspace{-12pt}
    \label{fig:overview}
\end{figure*}

\section{Proposed Approach}

\subsection{Overview}

\noindent
\textbf{Problem Formulation.}
Figure~\ref{fig:overview}(a) illustrates the overview of \textsc{Gamora}.
The inputs are flattened gate-level netlists in AIG format, without any micro-architectural or RTL information.
These AIGs are generated by the logic synthesis tool ABC~\cite{brayton2010abc}.
The goal is to exploit a multi-task GNN to reason high-level abstractions by performing node-level classification on AIGs, after which functional blocks (e.g., adders) can be extracted based on the annotated AIGs.

\noindent
\textbf{Case Study on Multipliers.}
Integer multipliers are indispensable to computationally intensive applications, such as signal processing and cryptography applications.
Recent years also witness the strong demand for large integer multipliers in homomorphic encryption~\cite{acar2018survey}.
In general, formal multiplier verification is challenging, especially for structurally complex designs such as Booth multipliers ~\cite{ciesielski2019understanding,mahzoon2019revsca,temel2021sound}.
Symbolic computer algebra (SCA) has been successfully employed to verify a variety of integer multipliers~\cite{yu2017fast,ciesielski2019understanding,mahzoon2019revsca,kaufmann2019verifying,alireza2022formal}, which relies heavily on detecting full adders (FAs) and half adders (HAs) in multiplier netlists.
The state-of-the-art implementation in ABC framework~\cite{yu2017fast} develops a fast algebraic rewriting approach to extracting adder trees from flattened multiplier netlists by detecting pairs of XOR and MAJ functions, which can handle large bitwidth multipliers (up to 2048-bit) but with extremely long runtime.
Thus, targeting integer multipliers, we leverage GNNs to identify XOR and MAJ functions to extract adders from flattened netlists, which improves the efficiency of word-level abstraction from BNs and has strong scalability enabled by GPU acceleration.

\subsection{Multi-Task Learning for Boolean Reasoning}

Boolean reasoning requires gathering structural and functional information from neighbor nodes, a process that can be imitated by the message-passing mechanism in GNNs.
The task of reasoning high-level abstractions from flattened netlists, i.e., pinpointing adders from AIGs, involves a two-step procedure~\cite{li2013wordrev,subramanyan2013reverse,yu2017fast}: (1) detecting XOR/MAJ functions to construct adders, and then (2) identifying their boundaries. 
Therefore, we propose to apply multi-task learning (MTL) for Boolean reasoning to approach its nature, in which knowledge sharing across sub-tasks provides higher reasoning precision.
This section details (1) how structural and functional information are fused in node embeddings, (2) how the two-step reasoning is formulated as a multi-task node classification, and (3) the post-processing after performing reasoning on each node in AIGs.

\subsubsection{Fusing Structural and Functional Information}
We leverage the message propagation and neighborhood aggregation in GNNs to generate the node embeddings of AIGs that simultaneously fuse structural and functional information.
First, the structural information is distilled by passing node embeddings along edges that connect them.
Second, the Boolean functional information can be encoded in node features.
For each node, there are three node features represented in binary values denoting node types and Boolean functionality.
The first node feature indicates whether this node is a PI/PO or intermediate node (i.e., AND gate).
The second and the third node features indicate whether each input edge is inverted or not, such that AIGs can be represented as homogeneous graphs without additional edge features.
These compressed node features not only encapsulate Boolean functionality of each node but also enable high compute and memory efficiency.
Figure~\ref{fig:overview}(b) shows the AIG of a 3-bit CSA multiplier, in which the structural information is presented in the AIG topology and the functional information is encoded in node features.
For example, node 1 is a PI with the feature vector [0, 0, 0]; node 7 is an internal node without negation on inputs, so the feature vector is [1, 0, 0]; node 17 has two inputs inverted, with the feature vector [1, 1, 1].

With the emphasis on generalization from simple to complex designs, the specific model employed is GraphSAGE~\cite{hamilton2017inductive}.
Given a GraphSAGE model with $K$ layers, the node embeddings propagated between different layers are computed as follows:

\begin{equation}
\begin{split}
    & h^k_{\mathcal{N}(v)} \gets \textsc{aggregate}_k(\{h_u^{k-1}, \forall u \in \mathcal{N}(v)\}); \\
    & h_v^k \gets \sigma(\textbf{W}^k \cdot \textsc{concat}(h^{k-1}_v, h^k_{\mathcal{N}(v)})).
\end{split}
\label{eq:message}
\end{equation}

\noindent
Here, $\mathcal{N}(v)$ is the immediate neighborhood of node $v$;
$\textsc{aggregate}_k$ and $\textbf{W}^k$ are the aggregation function and the weight matrix for layer $k$, respectively, where $\forall k \in \{1, ..., K\}$.
After stacking $K$ layers, the structural and functional information within $K$-hop search depth is fused in the embedding of each node.

\subsubsection{Multi-Task Classification}
We identify the Boolean function of each node by multi-task node classification to approach the nature of the problem: there are two steps involved in reasoning functional blocks from unstructured AIGs.
The first step detects XOR and MAJ functions from AIGs, which will be used to construct adders.
Since each XOR/MAJ function consists of multiple nodes in AIGs, only the root nodes of these functions are labeled as XOR/MAJ with other nodes marked as plain nodes.
In addition to the exact XOR/MAJ functions, negation-permutation-negation equivalent functions are also labeled as XOR/MAJ.
The second step aims to automatically identify the boundaries of HAs and FAs, and thus we label roots (i.e. the sum and the carry-out functions) and leaves of each adder.
Figure~\ref{fig:overview}(c) shows a multi-label annotated AIG of a 3-bit multiplier, using the ground truth provided by ABC.
Notably, one node can have multiple labels.
For example, node 20 is labeled with XOR and the root of an adder; node 17 is labeled with XOR.

The MTL not only follows the intuition of this two-step reasoning but also exploits divide and conquer, since it is extremely hard for GNNs to reach high prediction accuracy with a single-task multi-label node classification.
The employment of MTL enables knowledge sharing across sub-tasks and improves sample efficiency during training, which guarantees high reasoning performance.
Specifically, the two-step reasoning is decoupled into three simpler classification tasks using generated node embeddings:
\textit{Task 1} classifies the roots and leaves of adders;
\textit{Task 2} and \textit{Task 3} detect XOR and MAJ nodes, respectively.
We use hard parameter sharing for MTL and the overall loss function $\mathcal{L}$ is shown below:
\begin{equation}
\label{eq:loss}
    \mathcal{L} = \alpha \cdot \ell (\hat{y_1}, y_1) + \beta \cdot \ell(\hat{y_2}, y_2) + \gamma \cdot \ell(\hat{y_3}, y_3),
\end{equation}
in which $\ell$ is the negative log-likelihood between predictions (i.e., $\hat{y_1}$, $\hat{y_2}$, and $\hat{y_3}$) and the ground truth (i.e., $y_1$, $y_2$, and $y_3$), and $\alpha$, $\beta$, and $\gamma$ are hyper-parameters to adjust the importance of each task.
In our implementation, $\alpha = 0.8$ and $\beta = \gamma = 1$.

\subsubsection{Adder Tree Extraction from Multi-Labeled Graphs}
After performing the multi-task node classification, we can recognize XOR, MAJ, and root nodes of adders.
The XOR and MAJ pairs with identical inputs are matched to construct adders.
The conversion from Figure~\ref{fig:overview}(c) to \ref{fig:overview}(d) depicts the adder tree extraction.
In Figure~\ref{fig:overview}(c), the AIG has a set of XOR nodes $\mathbb{X}=$ \{12, 17, 20, 24, 29, 33, 36, 41, 44\} and a set of MAJ nodes $\mathbb{M}=$ \{10, 22, 25, 27, 37, 45\}.
After removing the nodes that are not marked as adder roots, $\mathbb{X}=$ \{12, 20, 24, 29, 36, 44\}.
Given $\mathbb{X}$ and $\mathbb{M}$,
node 12 is XOR3(8, 9, 0) and node 10 is MAJ3(8, 9, 0), a three-input XOR/MAJ function with node 8, node 9, and the constant zero as the inputs;
node 20 is XOR3(10, 13, 14) and node 15 is MAJ3(10, 13, 14);
this matching process continues until all six pairs of XOR and MAJ are generated, which are three FAs and three HAs, as shown in Figure~\ref{fig:overview}(d).

Notably, \textsc{Gamora} adopts graph learning to mimic the \textit{exact} reasoning.
In Figure~\ref{fig:overview}(e), one HA cannot be automatically extracted due to the misprediction of node 10. 
Our evaluation indicates only several nodes near the least significant bit are always mispredicted due to their shallow neighborhood structure, which has a subtle impact on the efficiency of algebraic rewriting.
By fusing structural and functional information into node embeddings and using MTL to approach the reasoning nature, 
\textsc{Gamora} is expected to reach as close as possible to the \textit{exact} reasoning precision.

\section{Experiment}
\vspace{-3pt}
\subsection{Experiment Setup}

The AIG-based CSA and Booth multipliers are generated by the logic synthesis tool ABC~\cite{brayton2010abc}, with the ground truth provided by the adder tree extraction command~\cite{yu2017fast}.
We consider two technology libraries: (1) the reduced standard-cell library \textit{mcnc.genlib} (with gate input size <=3) from SIS distribution~\cite{sentovich1992sis}, and (2) ASAP 7nm technologies \cite{xu2017standard}.
The GNN-based framework is implemented in Pytorch Geometric~\cite{Fey/Lenssen/2019}.
Two GraphSAGE models are developed for simple and complex design netlists: (1) a shallow 4-layer model with the hidden channel of 32 (for CSA multipliers w/ and w/o simple technology mapping), and (2) a deep 8-layer model with the hidden channel of 80 (for Booth multipliers and after complex technology mapping).
The generated node embeddings are passed to a shared linear layer with size of 32 and the ReLU activation function, followed by another linear transformation with softmax for each sub-task to perform node classification.
Experiments are performed on a Linux host with AMD EPYC 7742 64-core CPUs and one NVIDIA A100 SXM 40GB GPU.
\textit{
In general, \textsc{Gamora} is trained on small bitwidth multipliers (typically less than 32-bit) and evaluated on large bitwidth multipliers (up to 2048-bit).}



\begin{figure*}[ht]
    \centering
    \includegraphics[width=\textwidth]{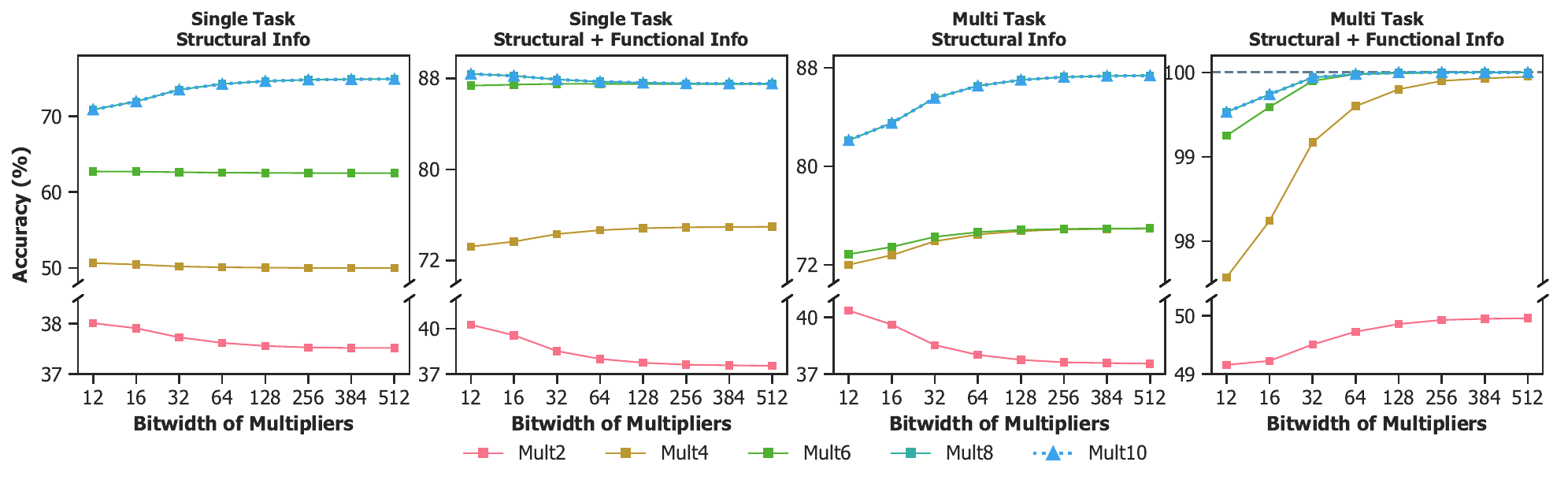}
    \vspace{-20pt}
    \caption{Sensitivity analysis on CSA multipliers with respect to (1) the bitwidth of multipliers for training (ranging from 2-bit to 10-bit), (2) single/multi-task, and (3) whether employing functional information.}
    \vspace{-12pt}
    \label{fig:sensitivity}
\end{figure*}

\begin{figure*}[h]
    \centering
    \includegraphics[width=\textwidth]{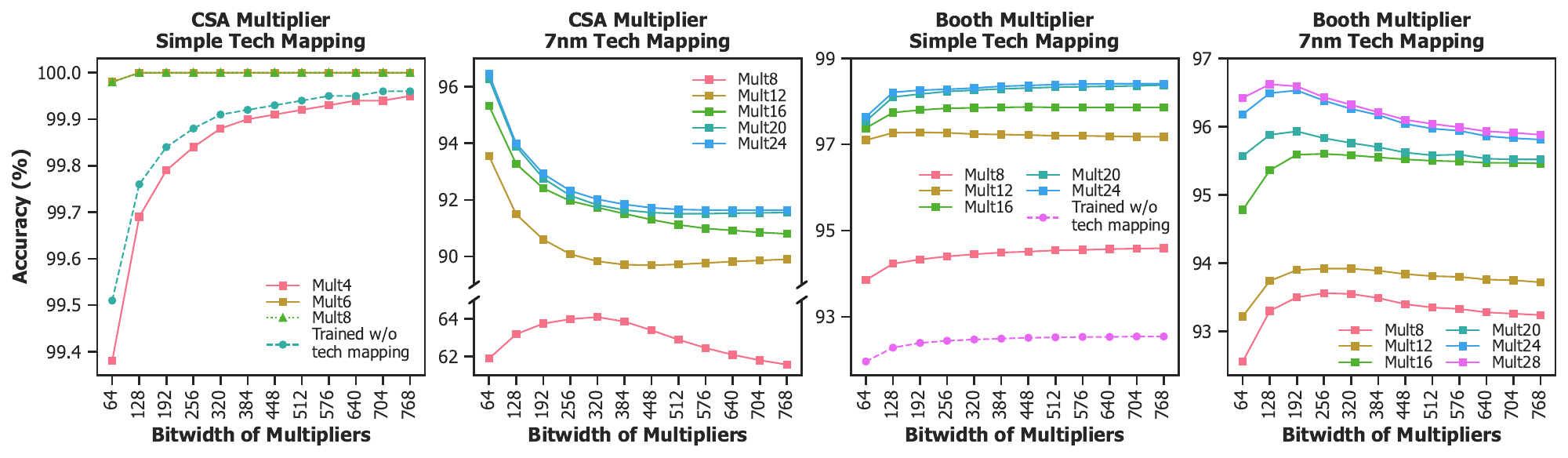}
     \vspace{-15pt}
    \caption{Evaluation on CSA and Booth multipliers, with simple and complex technology mapping.}
    \vspace{-10pt}
    \label{fig:techmapping}
\end{figure*}

\begin{figure}[h]
    \centering
    \includegraphics[width=\linewidth]{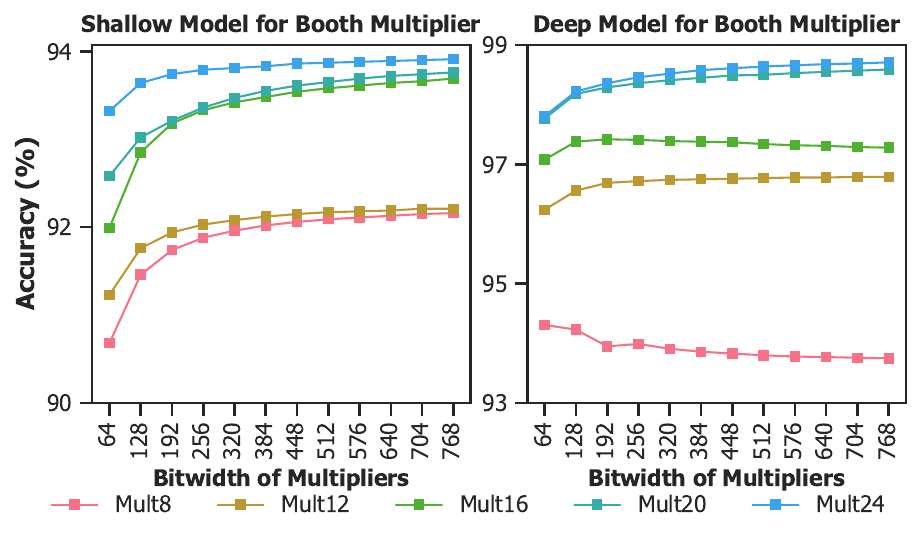}
    \vspace{-20pt}
    \caption{Evaluation on Booth multipliers with shallow and deep models.}
    \vspace{-15pt}
    \label{fig:booth}
\end{figure}

\vspace{-5pt}
\subsection{Evaluation on Reasoning Performance}

We evaluate the reasoning performance from three aspects:
(1) how functional and structural information influence the reasoning precision;
(2) how design complexity affects model selection and training;
(3) how technology mapping complicates the reasoning process and what domain insights can be derived to facilitate more accurate symbolic reasoning on complex BNs.

\vspace{0pt}
\subsubsection{Reasoning Precision Analysis}
Figure~\ref{fig:sensitivity} illustrates how the reasoning performance on CSA multipliers is affected by different bitwidth multipliers for training, single/multi-task setting, and the employment of functional information.
First, the larger bitwidth multiplier is adopted for training, the higher reasoning precision can be achieved, which typically converges after training with 8-bit multipliers.
The main reason is for CSA multipliers, an 8-bit multiplier is able to provide a sufficient variety of structural properties, which can be learned and well generalized to larger multipliers by \textsc{Gamora}.
Second, the multi-task setting conspicuously outperforms the single-task counterpart, indicating that the knowledge sharing across multiple tasks greatly benefits the prediction accuracy of every single task.
Third, there is always a boost of accuracy when employing functional information for prediction, since identifying the role of each node relies on not only the surrounding structure but also the function of itself and its neighbors.
The synergy of structural and functional information in \textsc{Gamora} is analogous to the combination of structural hashing and functional propagation in conventional symbolic reasoning.

With the multi-task setting and simultaneously fusing structural and functional attributes, \textsc{Gamora} achieves almost 100\% prediction accuracy in symbolic reasoning for CSA multipliers.
It is noted that several nodes near the least significant bit (LSB) are always mispredicted due to their shallow neighborhood structure, as shown in Figure~\ref{fig:overview}(e).
This means the HA at LSB cannot be automatically extracted, but can be easily corrected during post-processing.

\vspace{0pt}
\subsubsection{The Impact of Design Complexity}
We analyze the impact from design complexity by evaluating the reasoning performance on radix-4 Booth-encoded multipliers, as shown in Figure~\ref{fig:booth}.
From the model selection aspect, as Booth multipliers generally have more complex structures, deeper models are necessary to characterize neighborhood structures and provide informative node embeddings, thus guaranteeing high prediction accuracy.
From the training aspect, larger multipliers (i.e., up to 24-bit Booth multiplier) are required for training such that adequate variety and representativeness of structural and functional characteristics are exposed to and well captured by \textsc{Gamora}.

\vspace{0pt}
\subsubsection{The Impact of Technology Mapping}
It is a known challenge that technology mapping can increase the complexity of formal reasoning on BNs~\cite{li2013wordrev,subramanyan2013reverse,yu2016formal}. 
Thus, we evaluate the performance of \textsc{Gamora} with respect to different technology mapping options. The multipliers are mapped using the ABC standard-cell mapper (command \texttt{map}).
Figure~\ref{fig:techmapping} depicts the reasoning performance on CSA and Booth multipliers after simple technology mapping~\cite{sentovich1992sis} and ASAP 7nm technology mapping~\cite{xu2017standard}. 
Specifically, the ASAP 7nm library contains 161 standard-cell gates, including multi-output cells such as the full adder cell, which significantly increases the complexity and irregularity of post-mapping netlists.

In the simple technology mapping case, the models trained before technology mapping demonstrate good generalization capability, still reaching over 99\% and 92\% prediction accuracy for CSA and Booth multipliers, respectively;
with retraining, comparable reasoning performance to those on original multipliers is achieved with similar sizes of training multipliers.
The scenario is fairly different in the case of ASAP 7nm technology mapping, which employs a relatively complex technology library:
first, the generalization capability is limited before and after technology mapping;
second, the prediction accuracy slightly drops even with retraining;
third, it is necessary to use large training multipliers to guarantee performance.

These observations imply several takeaways.
First, the more complex technology library is applied, the more difficult it is for learning-based symbolic reasoning, since more complexity is involved both in AIG structures and the functionality of each node. This also implicates attributes related to the technology library should be included in node and edge features.
Second, the capability to cope with intricate AIG netlists comes at the expense of more comprehensive training data.
One underlying assumption of many supervised ML tasks is the training and testing data should be independent and identically distributed, which is governed by a fundamental principle called empirical risk minimization that provides theoretical performance bounds~\cite{vapnik1991principles}.
Thus, increasing the size of training data can envelop more knowledge of interested statistical properties, ensuring better generalization to testing data.

\subsection{Runtime and Scalability Analysis}
In addition to the high reasoning performance, we demonstrate the superiority of \textsc{Gamora} by analyzing its runtime and scalability.

\begin{figure}[t]
    \centering
    \includegraphics[width=\linewidth]{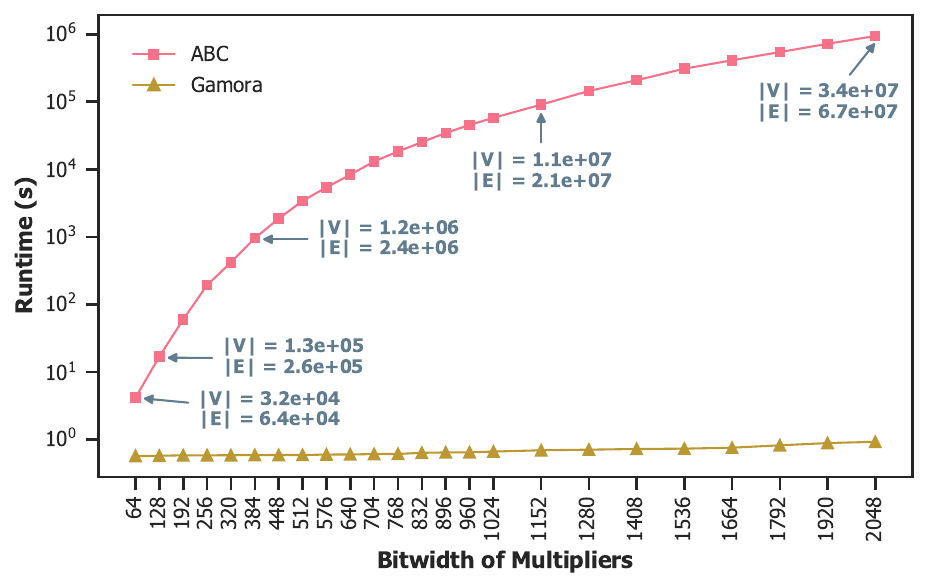}
    \vspace{-20pt}
    \caption{Runtime comparison between \textsc{Gamora} and ABC. Note that the number of nodes $|V|$ and the number of edges $|E|$ are annotated for scalability analysis.}
    \label{fig:runtime}
\end{figure}

\begin{figure}[t]
    \centering
    \vspace{-10pt}
    \includegraphics[width=1.0\linewidth]{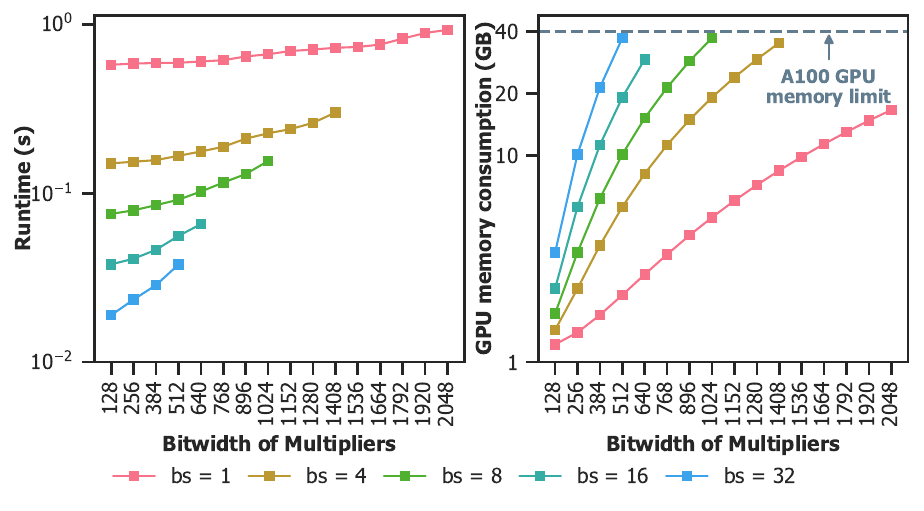}
    \vspace{-20pt}
    \caption{Average runtime and GPU memory consumption with batched reasoning, where the batch size is denoted as bs. We currently focus on single-GPU implementation.}
    \vspace{-10pt}
    \label{fig:batch}
\end{figure}

\noindent
\textbf{Runtime complexity analysis.}
Basically, the runtime only relates to the scale of AIGs, i.e., the number of nodes $|V|$ and the number of edges $|E|$.
Figure~\ref{fig:runtime} compares the runtime of \textsc{Gamora} against ABC on CSA multipliers:
for large designs such as a 2048-bit CSA multiplier with around 34 million nodes and 67 million edges, \textsc{Gamora} attains a speedup of up to six orders of magnitude.
This shows not only the great efficiency in symbolic reasoning enabled by graph learning but also the scalability to extremely large designs.

\noindent
\textbf{Batched reasoning with single GPU.}
Figure~\ref{fig:batch} shows further acceleration allowed by batched reasoning.
Currently, we focus on single GPU implementation, which limits the batch size by the GPU memory, and leave multi-GPU implementation as our future work to support larger batch processing.
Even with a single GPU, there already reveal promising results and positive trends benefiting from parallel execution and GPU acceleration.

\section{Conclusion}

Reasoning high-level abstractions from bit-blasted BNs has benefited functional verification, logic minimization, datapath synthesis, malicious logic identification, etc.
In this work, we propose a novel symbolic reasoning framework, \textsc{Gamora}, which exploits GNNs to imitate structural hashing and functional aggregation in conventional reasoning approaches.
Evaluation shows that 
(1) with the proposed multi-task GNN model, \textsc{Gamora} offers \textbf{high reasoning performance} that reaches almost 100\% and over 97\% accuracy for CSA and Booth-encoded multipliers, which is still over 92\% in finding functional modules after complex technology mapping;
(2) with GPU acceleration on graph learning, \textsc{Gamora} has \textbf{strong scalability} to BNs with over 33 million nodes, with up to \textbf{six orders of magnitude speedups} compared to the state-of-the-art implementation in the ABC framework; 
(3) \textsc{Gamora} also demonstrates {\bf great generalization capability} from simple to complex designs, such as from small to large bitwidth multipliers, and from before to after technology mapping. 
\textsc{Gamora} reveals the great potential of applying GNNs and GPU acceleration to speed up symbolic reasoning.


\vspace{-5pt}
\section{Acknowledge}

\noindent
This work is supported by National Science Foundation (NSF) under NSF-2047176, NSF-2019336, NSF-2008144, and NSF-2229562 awards.
\vspace{-10pt}

\bibliographystyle{plain}
\bibliography{ref}

\end{document}